\documentclass[aps,pra,reprint,onecolumn,notitlepage,eqsecnum,showkeys]{revtex4-2}

\usepackage{hyperref}

\newcommand{\bq}{\begin{eqnarray}}
\newcommand{\eq}{\end{eqnarray}}
\newcommand{\xx}{{\bf x}}
\newcommand{\yy}{{\bf y}}
\newcommand{\rr}{{\bf r}}

\begin{document}
\title{Thermodynamic limit of the free electron gas on a circle}

\author{Riccardo Fantoni}
\email{riccardo.fantoni@scuola.istruzione.it}
\affiliation{Universit\`a di Trieste, Dipartimento di Fisica, strada
  Costiera 11, 34151 Grignano (Trieste), Italy}

\date{\today}

\begin{abstract}
We show that for the ground state of a one dimensional free electron gas on a circle the 
analytic expression for the canonical ensemble partition function can be easily derived from 
the density matrix by assuming that the thermodynamic limit coincides with the limit of the 
eigenfunction expansion of the kinetic energy. This approximation fails to give the finite 
temperature partition function because those two limits cannot be chosen as coincident.  
\end{abstract}


\maketitle

\section{Introduction}

In statistical physics textbooks, like for example the Feynman, R P (1972) ``Statistical 
Mechanics: A Set of Lectures'' \cite{FeynmanFIP} section 2.8, the derivation of an analytic 
expression 
for the partition function of the free fermion or boson gas is accomplished choosing to work 
in the grand canonical ensemble. In this brief paper we show the difficulties one goes 
through if he insists in choosing to work in the canonical ensemble instead. For definiteness 
we will consider polarized fermions.

Some recent studies on the electron gas or {\sl the jellium} are about two dimensional 
systems \cite{Fantoni95a,Fantoni95b,Fantoni03a,Fantoni08c,Fantoni12b,Fantoni18c,Fantoni19a,Fantoni23a} or three dimensional ones 
\cite{Fantoni95b,Fantoni13g,Fantoni16b,Fantoni18a,Fantoni21b,Fantoni21d,Fantoni21i}. Here we will just consider an 
ideal electron gas in one dimension at a finite absolute temperature $T$.

The main actor of our problem is the thermal density operator $\hat{\rho}=e^{-\beta \hat{H}}$ 
where $\hat{H}$ is the Hamiltonian operator and $\beta=1/k_BT$ with $k_B$ Boltzmann's constant. 
We will only work in position representation so that 
$\rho(\rr,\rr';\beta)=\langle \rr|e^{-\beta \hat{H}}|\rr'\rangle$.

\section{A simple derivation}

Consider first one single free electron of mass $m$ in a one dimensional box of width $L$ 
with periodic boundary conditions, which is the same as saying that the electron lives in a 
circle. Its wave function $\psi(x)$ is such that 
$\psi(x+L)=\psi(x)$ and satisfies Shr\"odinger's equation, namely
\bq \label{se}
-\lambda\frac{\partial^2\psi(x)}{\partial x^2}=E\psi(x),
\eq
where $\lambda=\hbar^2/(2m)$. 

The solution of Eq. (\ref{se}) is as follows \cite{Landau3}
\bq
E_n&=&\lambda\left(\frac{2\pi}{L}\right)^2n^2,~~~n=0,1,2,3,\ldots\\
\psi_n(x)&=&\frac{1}{\sqrt{L}}\exp\left(i\frac{2\pi}{L}nx\right),~~~0<x<L
\eq
where $E_n$ are the eigenvalues and $\psi_n$ the normalized eigenvectors.

At an inverse temperature $\beta=1/k_BT$, the exact density matrix 
$\rho_1(x,y;\beta)=\sum_{n=-\infty}^\infty\psi_n^\star(x)\psi_n(y)\exp(-\beta E_n)$ for one
of those fermions in periodic boundary conditions is,
\bq \nonumber
\rho_1(x,y;\beta)&=&\frac{1}{L}\theta_3\left(\frac{\pi}{L}(x-y),\exp\left(
-\beta \lambda \left(\frac{2\pi}{L}\right)^2\right)\right)\\ \nonumber
&=&\lim_{q\rightarrow\infty}\frac{1}{L}\sum_{n=-q}^q\exp\left(-\beta\lambda
\left(\frac{2\pi}{L}\right)^2n^2\right)\exp\left(-i\frac{2\pi}{L}n(x-y)\right)\\ \label{rho1}
&=&\lim_{q\rightarrow\infty} k_q(x,y;\beta)~~,
\eq  
where $\theta_3(z,q)$ is a theta function (see Abramowitz and Stegun (1964) \cite{AbrSte}, 
Chapter 16, for its properties).  

Consider now $N=2p+1$ (with $p=0,1,2,3,\ldots$) free polarized fermions on a 
circle of circumference $L$. Usually for an electron gas it is more common to introduce 
Hartree's units where lengths are given in units of $a=L/N=1/\rho$, with $\rho$ the density of 
the gas, energies are given in Rydbergs $\hbar^2/(2ma_0^2)$, where $a_0=\hbar^2/(me^2)$, with 
$e$ the electron charge, is Bohr's radius. And the kinetic energy scales like $1/r_s^2$ with 
$r_s=a/a_0$. But since here we are dealing with a non interacting gas, we prefer not to use 
these conventions which would only make formulas less intuitive and pedagogic.

The density matrix of the $N$ fermions is now \cite{Landau3,FeynmanFIP},
\bq \nonumber
\rho_N(\xx,\yy;\beta)&=&\frac{1}{N!}\det\{\rho_1(x_i,y_j;\beta)\}_{i,j=1}^N
\\ \nonumber
&=&\lim_{q\rightarrow\infty}\frac{1}{N!}\det\{k_q(x_i,y_j;\beta)\}_{i,j=1}^N
\\ \label{dm}
&=&\lim_{q\rightarrow\infty}K_q(\xx,\yy;\beta)~~,
\eq
where $\xx=(x_1,x_2,\ldots,x_N)$, $\yy=(y_1,y_2,\ldots,y_N)$,
and $y_i$, $x_j$ are the initial and final positions of the $N$
fermions.

Notice that because of Pauli' s principle \cite{Landau3} (see appendix \ref{app}),
\bq
K_q=0~~~\mbox{when}~~~q<p~~.
\eq
For the particular case $q=p$ there is a simple expression for 
$K_q$, namely,
\bq \nonumber
K_p(\xx,\yy;\beta)=\frac{1}{N!}\frac{2^{N(N-1)}}{L^N}\exp\left(-2\beta\lambda
\left(\frac{2\pi}{L}\right)^2\sum_{n=1}^pn^2\right)\\ \label{ground}
\prod_{1\le i < j \le N}\sin\left(\frac{\pi}{L}(x_i-x_j)\right)
\sin\left(\frac{\pi}{L}(y_i-y_j)\right)~~.
\eq
This expression is the exact density matrix of the ground state
(when $\beta \rightarrow \infty$) of 
the $N$ fermions.

For example let's find the partition function 
$Z(\beta)={\rm tr}(\hat{\rho}_N)=\int\rho_N(\xx,\xx;\beta)\,d\xx$ of the 
fermion system in the thermodynamic limit. 
We need to calculate the trace $Z_p(\beta)$ of 
$K_p(\xx,\yy;\beta)$ and then take $p$ to infinity.
\bq \nonumber
Z_p(\beta)&=&\int_{-L/2}^{L/2}dx_1\,\cdots \int_{-L/2}^{L/2}dx_N\, 
K_p(\xx,\xx;\beta)\\
&=& \exp\left(-2\beta\lambda\left(\frac{2\pi}{L}\right)^2\sum_{n=1}^pn^2\right)
\frac{1}{N!} \frac{2^{N(N-1)}}{(2\pi)^N}I_N~~,
\eq
where,
\bq \nonumber
I_N&=&\int_{-\pi}^{\pi}d\theta_1\,\cdots \int_{-\pi}^{\pi}d\theta_N\,
\prod_{1\le i < j \le N}\sin^2((\theta_i-\theta_j)/2)\\
&=& N!  \frac{(2\pi)^N}{2^{N(N-1)}}~~.
\eq
So we get,
\bq
Z_p(\beta)=\exp\left(-2\beta\lambda\left(\frac{2\pi}{L}\right)^2\sum_{n=1}^pn^2\right)~~.
\eq
Or for the Helmholtz free energy, $F=-\ln Z/\beta$,
\bq \nonumber
F_p(\beta)&=&2\lambda\left(\frac{2\pi}{L}\right)^2\sum_{n=1}^pn^2\\
&=& \frac{\pi^2}{3}\rho^2\lambda\frac{N^2-1}{N}~~.
\eq
And in the thermodynamic limit,
\bq
f(\beta)=\lim_{p\rightarrow\infty}F_p(\beta)/N=\frac{\pi^2}{3}\rho^2
\lambda~~.
\eq
As expected the free energy is independent of temperature in the 
thermodynamic limit. Moreover we found the expected results for the
ground state energy
\bq
E_0=\lambda L\int_{-k_F}^{k_F}k^2\,\frac{dk}{2\pi}=
\left(\frac{L}{2\pi}\right)\frac{2}{3}\lambda k_F^3=
N\left(\frac{\lambda\rho^2\pi^2}{3}\right),
\eq
where the Fermi wave vector is $k_F=\pi\rho$.

But we see from equation (\ref{dm}) that in the thermodynamic limit 
(i.e. $p\rightarrow \infty$ and $\rho=N/L$ constant) it fails to give the exact density 
matrix of the fermions at finite inverse temperature $\beta$ for which it is necessary to 
relax the constraint $q=p$ and respect the order of the two limits, first the one over $q$ 
and only later the one over $p$.

\section{Conclusions}

When writing the canonical partition function of a free electron gas on a circle in the 
thermodynamic limit one has to deal with two kinds of limits: The limit of the eigenfunction 
expansion of the kinetic energy and the thermodynamic limit. In this brief paper we showed 
that if one takes the two limits as coincident then necessarily falls in the ground state 
case, the $\beta\to\infty$ limit. In this case in fact the zero temperature limit permits to 
take those two limits as the same. But in order to find the correct finite temperature 
case it is necessary to take those two limits independently in the correct order.

\section*{Conflict of Interests}
The Author has no conflict to disclose.

\appendix
\section{A determinantal identity}
\label{app}

Given three functions of two variables, K(x,y), L(x,y) and M(x,y)
such that,
\bq
K(x,y)=\sum_{n=-\infty}^{\infty}L(x,n)M(n,y)~~.
\eq
Take the following product,
\bq \nonumber
\lefteqn{K(x_1,y_{\pi 1})K(x_2,y_{\pi 2})\cdots K(x_n,y_{\pi n})=}
\\ \nonumber
&&\sum_{k_1,k_2,\ldots,k_n}[L(x_1,k_1)L(x_2,k_2)\cdots L(x_n,k_n)] \\
&&~~~~~~~~~~~[M(k_1,y_{\pi 1})M(k_2,y_{\pi 2})\cdots M(k_n,y_{\pi n})]~~.
\eq
Summing appropriately with respect to all permutations we obtain,
\bq \nonumber
\lefteqn{\det\{K(x_i,y_j)\}_{i,j=1}^n=}\\ \label{app1}
&&\sum_{k_1,k_2,\ldots,k_n}L(x_1,k_1)L(x_2,k_2)\cdots L(x_n,k_n)
\det\{M(k_i,y_j)\}_{i,j=1}^n~~.
\eq
The region of summation can be decomposed in nonoverlapping regions
$\Delta_\nu$ characterized by the inequalities $k_{\nu 1}<k_{\nu 2}<
\cdots <k_{\nu n}$, where $\nu$ is an arbitrary permutation of the set
$(1,2,\ldots ,n)$ into itself.

Transforming the region $\Delta_\nu$ by the change of variable
$k_{\nu i}\rightarrow k_i$ $(i=1,2,\ldots,n)$ and collecting the 
resulting sums, we obtain, for the righthand side of 
(\ref{app1}),
\bq \nonumber
\sum_{k_1<k_2<\ldots<k_n} \sum_\nu (-)^{|\nu|}
L(x_1,k_{\nu^{-1}1})L(x_2,k_{\nu^{-1}2})\cdots L(x_n,k_{\nu^{-1}n})
~~~~~~~~~~~~~~~\\ 
\det\{M(k_i,y_j)\}_{i,j=1}^n~~,
\eq  
where the signature $(-)^{|\nu|}$ in each term appears as a consequence
of rearranging the rows of $\det{M}$.

So we derived the following composition formula
\footnote{Which holds also after replacing the sums with integrals.},
\bq
\det\{K(x_i,y_j)\}_{i,j=1}^n=\sum_{k_1<k_2<\ldots<k_n}
\det\{L(x_i,k_j)\}_{i,j=1}^n \det\{M(k_i,y_j)\}_{i,j=1}^n~~.
\eq 

Applied to the function $k_q$ defined in (\ref{rho1}) as,
\bq
k_q(\theta,\phi)=\sum_{n=-q}^q\mu_n e^{in\theta}e^{-in\phi}~~,
\eq
we see that for $q\ge (N-1)/2$,
\bq \nonumber
\lefteqn{\det\{k_q(\theta_i,\phi_j)\}_{i,j=1}^N=}\\
&&\mu_0\prod_{n=1}^q|\mu_n|^2
\sum_{-q\le k_1<k_2<\ldots<k_n\le q}
\det\{e^{ik_j\theta_i}\}_{i,j=1}^N \det\{e^{-ik_i\phi_j}\}_{i,j=1}^N
~~.
\eq
So when $q=(N-1)/2$ the sum has only one term which is given by
equation (\ref{ground}). And for $q<(N-1)/2$, $\det\{k_q\}=0$.

\bibliography{1degf}

\begin{thebibliography}{18}%
\makeatletter
\providecommand \@ifxundefined [1]{%
 \@ifx{#1\undefined}
}%
\providecommand \@ifnum [1]{%
 \ifnum #1\expandafter \@firstoftwo
 \else \expandafter \@secondoftwo
 \fi
}%
\providecommand \@ifx [1]{%
 \ifx #1\expandafter \@firstoftwo
 \else \expandafter \@secondoftwo
 \fi
}%
\providecommand \natexlab [1]{#1}%
\providecommand \enquote  [1]{``#1''}%
\providecommand \bibnamefont  [1]{#1}%
\providecommand \bibfnamefont [1]{#1}%
\providecommand \citenamefont [1]{#1}%
\providecommand \href@noop [0]{\@secondoftwo}%
\providecommand \href [0]{\begingroup \@sanitize@url \@href}%
\providecommand \@href[1]{\@@startlink{#1}\@@href}%
\providecommand \@@href[1]{\endgroup#1\@@endlink}%
\providecommand \@sanitize@url [0]{\catcode `\\12\catcode `\$12\catcode
  `\&12\catcode `\#12\catcode `\^12\catcode `\_12\catcode `\%12\relax}%
\providecommand \@@startlink[1]{}%
\providecommand \@@endlink[0]{}%
\providecommand \url  [0]{\begingroup\@sanitize@url \@url }%
\providecommand \@url [1]{\endgroup\@href {#1}{\urlprefix }}%
\providecommand \urlprefix  [0]{URL }%
\providecommand \Eprint [0]{\href }%
\providecommand \doibase [0]{https://doi.org/}%
\providecommand \selectlanguage [0]{\@gobble}%
\providecommand \bibinfo  [0]{\@secondoftwo}%
\providecommand \bibfield  [0]{\@secondoftwo}%
\providecommand \translation [1]{[#1]}%
\providecommand \BibitemOpen [0]{}%
\providecommand \bibitemStop [0]{}%
\providecommand \bibitemNoStop [0]{.\EOS\space}%
\providecommand \EOS [0]{\spacefactor3000\relax}%
\providecommand \BibitemShut  [1]{\csname bibitem#1\endcsname}%
\let\auto@bib@innerbib\@empty
\bibitem [{\citenamefont {Feynman}(1972)}]{FeynmanFIP}%
  \BibitemOpen
  \bibfield  {author} {\bibinfo {author} {\bibfnamefont {R.~P.}\ \bibnamefont
  {Feynman}},\ }\href@noop {} {\emph {\bibinfo {title} {{Statistical Mechanics:
  A Set of Lectures}}}},\ \bibinfo {series} {Frontiers in Physics},
  Vol.~\bibinfo {volume} {36}\ (\bibinfo  {publisher} {W. A. Benjamin, Inc.},\
  \bibinfo {year} {1972})\ \bibinfo {note} {notes taken by R. Kikuchi and H. A.
  Feiveson, edited by Jacob Shaham}\BibitemShut {NoStop}%
\bibitem [{\citenamefont {Fantoni}\ and\ \citenamefont
  {Tosi}(1995{\natexlab{a}})}]{Fantoni95a}%
  \BibitemOpen
  \bibfield  {author} {\bibinfo {author} {\bibfnamefont {R.}~\bibnamefont
  {Fantoni}}\ and\ \bibinfo {author} {\bibfnamefont {M.~P.}\ \bibnamefont
  {Tosi}},\ }\bibfield  {title} {\bibinfo {title} {Decay of correlations and
  related sum rules in a layered classical plasma},\ }\href
  {https://doi.org/10.1007/BF02451594} {\bibfield  {journal} {\bibinfo
  {journal} {Nuovo Cimento}\ }\textbf {\bibinfo {volume} {17D}},\ \bibinfo
  {pages} {155} (\bibinfo {year} {1995}{\natexlab{a}})}\BibitemShut {NoStop}%
\bibitem [{\citenamefont {Fantoni}\ and\ \citenamefont
  {Tosi}(1995{\natexlab{b}})}]{Fantoni95b}%
  \BibitemOpen
  \bibfield  {author} {\bibinfo {author} {\bibfnamefont {R.}~\bibnamefont
  {Fantoni}}\ and\ \bibinfo {author} {\bibfnamefont {M.~P.}\ \bibnamefont
  {Tosi}},\ }\bibfield  {title} {\bibinfo {title} {Coordinate space form of
  interacting reference response function of d-dimensional jellium},\ }\href
  {https://doi.org/10.1007/BF02454131} {\bibfield  {journal} {\bibinfo
  {journal} {Nuovo Cimento}\ }\textbf {\bibinfo {volume} {17D}},\ \bibinfo
  {pages} {1165} (\bibinfo {year} {1995}{\natexlab{b}})}\BibitemShut {NoStop}%
\bibitem [{\citenamefont {Fantoni}\ \emph {et~al.}(2003)\citenamefont
  {Fantoni}, \citenamefont {Jancovici},\ and\ \citenamefont
  {T\'ellez}}]{Fantoni03a}%
  \BibitemOpen
  \bibfield  {author} {\bibinfo {author} {\bibfnamefont {R.}~\bibnamefont
  {Fantoni}}, \bibinfo {author} {\bibfnamefont {B.}~\bibnamefont {Jancovici}},\
  and\ \bibinfo {author} {\bibfnamefont {G.}~\bibnamefont {T\'ellez}},\
  }\bibfield  {title} {\bibinfo {title} {Pressures for a one-component plasma
  on a pseudosphere},\ }\href {https://doi.org/10.1023/A:1023671419021}
  {\bibfield  {journal} {\bibinfo  {journal} {J. Stat. Phys.}\ }\textbf
  {\bibinfo {volume} {112}},\ \bibinfo {pages} {27} (\bibinfo {year}
  {2003})}\BibitemShut {NoStop}%
\bibitem [{\citenamefont {Fantoni}\ and\ \citenamefont
  {T\'ellez}(2008)}]{Fantoni08c}%
  \BibitemOpen
  \bibfield  {author} {\bibinfo {author} {\bibfnamefont {R.}~\bibnamefont
  {Fantoni}}\ and\ \bibinfo {author} {\bibfnamefont {G.}~\bibnamefont
  {T\'ellez}},\ }\bibfield  {title} {\bibinfo {title} {Two dimensional
  one-component plasma on a flamm's paraboloid},\ }\href
  {https://doi.org/10.1007/s10955-008-9616-x} {\bibfield  {journal} {\bibinfo
  {journal} {J. Stat. Phys.}\ }\textbf {\bibinfo {volume} {133}},\ \bibinfo
  {pages} {449} (\bibinfo {year} {2008})}\BibitemShut {NoStop}%
\bibitem [{\citenamefont {Fantoni}(2012)}]{Fantoni12b}%
  \BibitemOpen
  \bibfield  {author} {\bibinfo {author} {\bibfnamefont {R.}~\bibnamefont
  {Fantoni}},\ }\bibfield  {title} {\bibinfo {title} {Two component plasma in a
  flamm's paraboloid},\ }\href
  {https://doi.org/10.1088/1742-5468/2012/04/P04015} {\bibfield  {journal}
  {\bibinfo  {journal} {J. Stat. Mech.}\ ,\ \bibinfo {pages} {04015}} (\bibinfo
  {year} {2012})}\BibitemShut {NoStop}%
\bibitem [{\citenamefont {Fantoni}(2018{\natexlab{a}})}]{Fantoni18c}%
  \BibitemOpen
  \bibfield  {author} {\bibinfo {author} {\bibfnamefont {R.}~\bibnamefont
  {Fantoni}},\ }\bibfield  {title} {\bibinfo {title} {One-component fermion
  plasma on a sphere at finite temperature},\ }\href
  {https://doi.org/10.1142/S012918311850064X} {\bibfield  {journal} {\bibinfo
  {journal} {Int. J. Mod. Phys. C}\ }\textbf {\bibinfo {volume} {29}},\
  \bibinfo {pages} {1850064} (\bibinfo {year}
  {2018}{\natexlab{a}})}\BibitemShut {NoStop}%
\bibitem [{\citenamefont {Fantoni}(2019)}]{Fantoni19a}%
  \BibitemOpen
  \bibfield  {author} {\bibinfo {author} {\bibfnamefont {R.}~\bibnamefont
  {Fantoni}},\ }\bibfield  {title} {\bibinfo {title} {Plasma living in a curved
  surface at some special temperature},\ }\href
  {https://doi.org/10.1016/j.physa.2019.04.222} {\bibfield  {journal} {\bibinfo
   {journal} {Physica A}\ }\textbf {\bibinfo {volume} {177}},\ \bibinfo {pages}
  {524} (\bibinfo {year} {2019})}\BibitemShut {NoStop}%
\bibitem [{\citenamefont {Fantoni}(2023)}]{Fantoni23a}%
  \BibitemOpen
  \bibfield  {author} {\bibinfo {author} {\bibfnamefont {R.}~\bibnamefont
  {Fantoni}},\ }\bibfield  {title} {\bibinfo {title} {One-component fermion
  plasma on a sphere at finite temperature. the anisotropy in the paths
  conformations},\ }\href {https://doi.org/10.1088/1742-5468/aceb54} {\bibfield
   {journal} {\bibinfo  {journal} {J. Stat. Mech.}\ ,\ \bibinfo {pages}
  {083103}} (\bibinfo {year} {2023})}\BibitemShut {NoStop}%
\bibitem [{\citenamefont {Fantoni}(2013)}]{Fantoni13g}%
  \BibitemOpen
  \bibfield  {author} {\bibinfo {author} {\bibfnamefont {R.}~\bibnamefont
  {Fantoni}},\ }\bibfield  {title} {\bibinfo {title} {Radial distribution
  function in a diffusion monte carlo simulation of a fermion fluid between the
  ideal gas and the jellium model},\ }\href
  {https://doi.org/10.1140/epjb/e2013-40204-3} {\bibfield  {journal} {\bibinfo
  {journal} {Eur. Phys. J. B}\ }\textbf {\bibinfo {volume} {86}},\ \bibinfo
  {pages} {286} (\bibinfo {year} {2013})}\BibitemShut {NoStop}%
\bibitem [{\citenamefont {Alastuey}\ and\ \citenamefont
  {Fantoni}(2016)}]{Fantoni16b}%
  \BibitemOpen
  \bibfield  {author} {\bibinfo {author} {\bibfnamefont {A.}~\bibnamefont
  {Alastuey}}\ and\ \bibinfo {author} {\bibfnamefont {R.}~\bibnamefont
  {Fantoni}},\ }\bibfield  {title} {\bibinfo {title} {Fourth moment sum rule
  for the charge correlations of a two-component classical plasma},\ }\href
  {https://doi.org/10.1007/s10955-016-1512-1} {\bibfield  {journal} {\bibinfo
  {journal} {J. Stat. Phys.}\ }\textbf {\bibinfo {volume} {163}},\ \bibinfo
  {pages} {887} (\bibinfo {year} {2016})}\BibitemShut {NoStop}%
\bibitem [{\citenamefont {Fantoni}(2018{\natexlab{b}})}]{Fantoni18a}%
  \BibitemOpen
  \bibfield  {author} {\bibinfo {author} {\bibfnamefont {R.}~\bibnamefont
  {Fantoni}},\ }\bibfield  {title} {\bibinfo {title} {Two component
  boson-fermion plasma at finite temperature},\ }\href
  {https://doi.org/10.1142/S0129183118500286} {\bibfield  {journal} {\bibinfo
  {journal} {Int. J. Mod. Phys. C}\ }\textbf {\bibinfo {volume} {29}},\
  \bibinfo {pages} {1850028} (\bibinfo {year}
  {2018}{\natexlab{b}})}\BibitemShut {NoStop}%
\bibitem [{\citenamefont {Fantoni}(2021{\natexlab{a}})}]{Fantoni21b}%
  \BibitemOpen
  \bibfield  {author} {\bibinfo {author} {\bibfnamefont {R.}~\bibnamefont
  {Fantoni}},\ }\bibfield  {title} {\bibinfo {title} {Jellium at finite
  temperature using the restricted worm algorithm},\ }\href
  {https://doi.org/10.1140/epjb/s10051-021-00078-y} {\bibfield  {journal}
  {\bibinfo  {journal} {Eur. Phys. J. B}\ }\textbf {\bibinfo {volume} {94}},\
  \bibinfo {pages} {63} (\bibinfo {year} {2021}{\natexlab{a}})}\BibitemShut
  {NoStop}%
\bibitem [{\citenamefont {Fantoni}(2021{\natexlab{b}})}]{Fantoni21d}%
  \BibitemOpen
  \bibfield  {author} {\bibinfo {author} {\bibfnamefont {R.}~\bibnamefont
  {Fantoni}},\ }\bibfield  {title} {\bibinfo {title} {Form invariance of the
  moment sum-rules for jellium with the addition of short-range terms in the
  pair-potential},\ }\href {https://doi.org/10.1007/s12648-020-01750-2}
  {\bibfield  {journal} {\bibinfo  {journal} {Indian J. Phys.}\ }\textbf
  {\bibinfo {volume} {95}},\ \bibinfo {pages} {1027} (\bibinfo {year}
  {2021}{\natexlab{b}})}\BibitemShut {NoStop}%
\bibitem [{\citenamefont {Fantoni}(2021{\natexlab{c}})}]{Fantoni21i}%
  \BibitemOpen
  \bibfield  {author} {\bibinfo {author} {\bibfnamefont {R.}~\bibnamefont
  {Fantoni}},\ }\bibfield  {title} {\bibinfo {title} {Jellium at finite
  temperature},\ }\href {https://doi.org/10.1080/00268976.2021.1996648}
  {\bibfield  {journal} {\bibinfo  {journal} {Mol. Phys.}\ }\textbf {\bibinfo
  {volume} {120}},\ \bibinfo {pages} {4} (\bibinfo {year}
  {2021}{\natexlab{c}})}\BibitemShut {NoStop}%
\bibitem [{\citenamefont {Landau}\ and\ \citenamefont
  {Lifshitz}(1977)}]{Landau3}%
  \BibitemOpen
  \bibfield  {author} {\bibinfo {author} {\bibfnamefont {L.~D.}\ \bibnamefont
  {Landau}}\ and\ \bibinfo {author} {\bibfnamefont {E.~M.}\ \bibnamefont
  {Lifshitz}},\ }\href@noop {} {\emph {\bibinfo {title} {{Quantum Mechanics.
  Non-relativistic Theory.}}}},\ \bibinfo {edition} {3rd}\ ed.,\ Vol.~\bibinfo
  {volume} {3}\ (\bibinfo  {publisher} {Pergamon Press},\ \bibinfo {year}
  {1977})\ \bibinfo {note} {course of Theoretical Physics}\BibitemShut
  {NoStop}%
\bibitem [{\citenamefont {Abramowitz}\ and\ \citenamefont
  {Stegun}(1972)}]{AbrSte}%
  \BibitemOpen
  \bibfield  {author} {\bibinfo {author} {\bibfnamefont {M.}~\bibnamefont
  {Abramowitz}}\ and\ \bibinfo {author} {\bibfnamefont {I.~A.}\ \bibnamefont
  {Stegun}},\ }\href@noop {} {\emph {\bibinfo {title} {{Handbook of
  Mathematical Functions}}}}\ (\bibinfo  {publisher} {Dover Pubblications},\
  \bibinfo {year} {1972})\BibitemShut {NoStop}%
\bibitem [{Note1()}]{Note1}%
  \BibitemOpen
  \bibinfo {note} {Which holds also after replacing the sums with
  integrals.}\BibitemShut {Stop}%
\end{thebibliography}%

\end{document}